# MANIN TRIPLES AND N = 2 SUPERCONFORMAL FIELD THEORY

E. GETZLER

In this paper, we construct a family of WZW models with $N = 2$ superconformal symmetry, including a deformation of the Kazama-Suzuki model.

Recall that if $V$ is a complex vector space with a non-degenerate bilinear form, a polarization of $V$ is a splitting of $V$, $V = V_+ \oplus V_-$, into complementary isotropic subspaces. If the vector space is a Lie algebra $\mathfrak{g}$ and the inner product is invariant, it is natural to restrict attention to polarizations $\mathfrak{g} = \mathfrak{g}_+ \oplus \mathfrak{g}_-$ such that $\mathfrak{g}_\pm$ are isotropic Lie subalgebras of $\mathfrak{g}$; such a structure is called a Manin triple. Manin triples arise naturally in Drinfeld's approach to completely integrable systems and quantum groups [3].

As we explain in Section 1, Spindel, Sevrin, Troust and van Proyen ([10], [9]) have shown how to associate to a Manin triple an $N = 2$ superconformal field theory (the work of Kazama-Suzuki [7] is a special case of their results). We will construct a deformation of this theory, parametrized by an element $\alpha \in \mathfrak{g}_0$; here, $\mathfrak{g}_0$ is the subspace of $\mathfrak{g}$ orthogonal to $[\mathfrak{g}_+, \mathfrak{g}_+] \oplus [\mathfrak{g}_-, \mathfrak{g}_-] \subset \mathfrak{g}$. The $N = 2$ central charge of the deformed model is

$$d = \tfrac{1}{2} \dim \mathfrak{g} - (\rho, \rho) - (\alpha, \alpha),$$

where $\rho$ is a certain element of $\mathfrak{g}$, determined by the structure of a Manin triple on $\mathfrak{g}$.

In Section 2, we give an exposition of some formulas from the theory of Manin triples which we need. We show how to associate a Manin triple to a pair $(\mathbf{k}, \mathbf{p})$, where $\mathbf{k}$ is a simple Lie algebra and $\mathbf{p}$ is a parabolic subalgebra of $\mathbf{k}$. If $\mathbf{p} = \mathbf{b}$ is a Borel subalgebra, we obtain the Kazama-Suzuki model, while if $\mathbf{p} = \mathbf{k}$, we obtain the $G/G$ model. We will see that in the first case, any value of the central charge may be realized, while in the second case, $\rho$ and $\alpha$ lie in $\mathfrak{g}_+$, and thus $d = \dim \mathbf{k}$ is independent of the choice of $\alpha$.

In Section 3, we give the proof of $N = 2$ superconformal symmetry in our model.

In an appendix, we explain the formulas which are used in the manipulation of operator products: while straightforward, their application leads to very complicated formulas, and the chances of making an error are greatly reduced by the availability of an excellent Mathematica package (Thielemans [11]).

The author is partially supported by a fellowship of the Sloan Foundation, and a research grant of the NSF.

This work was inspired by the article of Parkhomenko [8], who drew attention to the central role played in [10] and [9] of Manin triples. We are also grateful to Bong Lian and Gregg Zuckerman for a number of interesting conversations.

## 1. BACKGROUND

Let us describe the results of Spindel et al. First recall the definition of the $N = 1$ superconformal algebra: this is the chiral algebra spanned by the stress-energy tensor $T(z)$ and a fermionic field $G(z)$, and their derivatives, with operator products

$$T(z) \cdot T(w) \sim \frac{\frac{3}{2}d}{(z-w)^4} + \frac{2T(w)}{(z-w)^2} + \frac{\partial T(z)}{z-w},$$

$$T(z) \cdot G(w) \sim \frac{\frac{3}{2}G(w)}{(z-w)^2} + \frac{\partial G(w)}{z-w},$$

$$G(z) \cdot G(w) \sim \frac{2d}{(z-w)^3} + \frac{2T(w)}{z-w}.$$

Here, $d$ is a real number called the central charge of the algebra. (This normalization of the central charge, related to the usual central charge $c$ by the formula $d = c/3$, is more natural in the study of $N = 2$ models.)

A representation of the $N = 1$ superconformal chiral algebra may be defined using a vector of scalar fields $\phi(z) = (\phi_1(z), \ldots, \phi_n(z))$, with operator products

$$\phi_i(z) \cdot \phi_j(w) \sim -\log(z-w)\delta_{ij},$$

and a vector of fermion fields $\psi(z) = (\psi_1(z), \ldots, \psi_n(z))$, with operator products

$$\psi_i(z) \cdot \psi_j(w) \sim \frac{\delta_{ij}}{z-w}.$$

Then given a vector $\alpha \in \mathbb{C}^n$, the fields

$$T = -\tfrac{1}{2}(\partial\phi, \partial\phi) + (\partial\psi, \psi) + i(\alpha, \partial^2\phi),$$
$$G = i(\partial\phi, \psi) + (\alpha, \partial\psi),$$

realize the $N = 1$ superconformal algebra, with central charge $d = n/2 - 4|\alpha|^2$.

The $N = 2$ superconformal algebra, an extension of the $N = 1$ superconformal algebra, is the chiral algebra spanned by the stress-energy tensor $T(z)$, a $U(1)$ current $J(z)$, a pair of fermionic fields $G^{\pm}(z)$, and their derivatives, satisfying the operator

products

$$G^+(z) \cdot G^-(w) \sim \frac{d}{(z-w)^3} + \frac{J(w)}{(z-w)^2} + \frac{T(w) + \frac{1}{2}\partial J(w)}{z-w},$$

$$G^\pm(z) \cdot G^\pm(w) \sim 0,$$

$$T(z) \cdot T(w) \sim \frac{\frac{3}{2}d}{(z-w)^4} + \frac{2T(w)}{(z-w)^2} + \frac{\partial T(z)}{z-w},$$

$$T(z) \cdot G^\pm(w) \sim \frac{\frac{3}{2}G^\pm(w)}{(z-w)^2} + \frac{\partial G^\pm(w)}{z-w},$$

$$T(z) \cdot J(w) \sim \frac{J(w)}{(z-w)^2} + \frac{\partial J(w)}{z-w},$$

$$J(z) \cdot J(w) \sim \frac{d}{(z-w)^2},$$

$$J(z) \cdot G^\pm(w) \sim \pm\frac{G^\pm(w)}{z-w}.$$

The stress-energy tensor $T$, together with either the field $G^+ + G^-$ or the field $(G^+ - G^-)/i$, span an $N = 1$ superconformal chiral algebra.

If $V$ is a complex vector space with a non-degenerate bilinear form, a polarization of $V$ is a splitting $V = V_+ \oplus V_-$ into complementary isotropic subspaces. If $V$ is the complexification of a real inner product space $V = W \otimes \mathbb{C}$ with positive definite inner product, then the spaces $V_\pm$ will be complex conjugate to each other, but if the inner product of $W$ is non-degenerate but not necessarily positive definite, there may be other types of polarization than such complex ones.

Given a polarization of the inner product space $\mathbb{C}^n$, we obtain an underlying $N = 2$ superconformal symmetry on the above $N = 1$ superconformal model (for a polarization to exist, $n$ must of course be even). Since all polarizations of $\mathbb{C}^n$ are equivalent, let us work with that in which

$$V_\pm = \text{span}\{x_{2j} \pm ix_{2j+1} \mid 1 \leq i \leq \ell = n/2\}.$$

Reassemble the real fields $\phi(z)$ and $\psi(z)$, and the vector $\alpha$, into complex fields $\Phi(z) = (\Phi_1, \ldots, \Phi_\ell)$ and $\Psi(z) = (\Psi_1, \ldots, \Psi_\ell)$, and a vector $\beta \in \mathbb{C}^\ell$, where $\ell = n/2$:

$$\Phi_j(z) = 2^{-1/2}(\phi_{2j}(z) + i\phi_{2j+1}(z)),$$
$$\Psi_j(z) = 2^{-1/2}(\psi_{2j}(z) + i\psi_{2j+1}(z)),$$
$$\beta_j = 2^{-1/2}(\alpha_{2j} + i\alpha_{2j+1}).$$

With these notations, the generators of the $N = 2$ chiral algebra are as follows:

$$T = -(\partial\bar{\Phi}, \partial\Phi) + \tfrac{1}{2}\big((\partial\bar{\Psi}, \Psi) - (\bar{\Psi}, \partial\Psi)\big) + \tfrac{i}{2}\big((\bar{\beta}, \partial^2\Phi) + i(\beta, \partial^2\bar{\Phi})\big),$$
$$G^+ = i(\partial\Phi, \bar{\Psi}) + (\beta, \partial\bar{\Psi}),$$
$$G^- = i(\partial\bar{\Phi}, \Psi) + (\bar{\beta}, \partial\Psi),$$
$$J = (\bar{\Psi}, \Psi) + i(\bar{\beta}, \partial\Phi) - i(\beta, \partial\bar{\Phi}).$$

In this paper, we generalize this model, replacing the vector space $\mathbb{R}^n$ by a reductive Lie algebra $\mathfrak{g}$.

Let $\mathfrak{g}$ be a reductive Lie algebra, and let $(-, -)$ be an invariant inner product on $\mathfrak{g}$. Let $x_i$ be a basis of $\mathfrak{g}$, relative to which $[x_i, x_j] = c_{ij}^k$ and $h_{ij} = (x_i, x_j)$. Denote $h_{kl}c_{ij}^l$ by $c_{ijk}$: it is antisymmetric in all three indices, reflecting the invariance of the inner product. Let $h^{ij}$ be the matrix inverse of $h_{ij}$, and let $\langle -, - \rangle$ be the Killing form of $\mathfrak{g}$, with coefficients

$$g_{ij} = \langle x_i, x_j \rangle = c_{ik}^l c_{jl}^k.$$

Let $\mathfrak{g}$ be a simple Lie algebra with Cartan subalgebra $\mathfrak{h}^*$ and highest root $\theta$. Let $h$ be the dual Coxeter number of $\mathbf{k}$. We say that an inner product $(-, -)$ has level $k$ if

$$\frac{(x, x)}{\langle x, x \rangle} = \frac{k + h}{2h};$$

the Killing form itself has level $h$.

Consider the chiral algebra generated by fields $J_i(z)$ satisfying the operator product expansions

$$J_i(z) \cdot J_j(w) = \frac{h_{ij} - \tfrac{1}{2}g_{ij}}{(z - w)^2} + \frac{c_{ij}^k J_k(w)}{z - w}.$$

These relations define a projective representation of the affine Kac-Moody algebra $\widehat{\mathfrak{g}}$, with central extension corresponding to the invariant inner product $(x, y) - \tfrac{1}{2}\langle x, y \rangle$ on $\mathfrak{g}$. Introduce free fermionic fields $\psi^i(z)$, with operator products

$$\psi^i(z) \cdot \psi^j(w) \sim \frac{h^{ij}}{z - w}.$$

Together with the currents $I_i = J_i - \tfrac{1}{2}c_{ijk}\psi^j\psi^k$, we obtain a realization of the super Kac-Moody algebra associated to $\mathfrak{g}$ (Kac-Todorov [6]),

$$I_i(z) \cdot I_j(w) \sim \frac{h_{ij}}{(z - w)^2} + \frac{c_{ij}^k I_k(w)}{z - w},$$
$$I_i(z) \cdot \psi^j(w) \sim -\frac{c_{ik}^j \psi^k(w)}{z - w};$$

thus, the fermionic fields lie in the coadjoint representation of $\mathfrak{g}$. Let $\alpha$ be an element of the Lie algebra $\mathfrak{g}$, and define the supersymmetry current $\mathbb{G}(z)$ by the formula

$$\mathbb{G} = J_i\psi^i - \tfrac{1}{6}c_{ijk}\psi^i\psi^j\psi^k + \alpha_i\partial\psi^i.$$

The field $\mathbb{G}(z)$ generates a realization of the $N = 1$ superconformal algebra, with stress-energy tensor
$$\mathbb{T} = \tfrac{1}{2}\big((J,J) + (\partial\psi,\psi) + (\alpha,\partial I)\big),$$
and central charge $d = \tfrac{1}{2}\dim\mathfrak{g} - \tfrac{1}{6}h_{ij}g^{ij} - (\alpha,\alpha)$. When $\mathfrak{g}$ is abelian, the Killing form vanishes, and this model becomes identical to the free-field realization of the $N = 1$ superconformal chiral algebra.

In Section 2, we will show that given the following data, this model has an underlying $N = 2$ superconformal symmetry:

(1) the structure of a Manin triple $(\mathfrak{g},\mathfrak{g}_+,\mathfrak{g}_-)$ on $\mathfrak{g}$;
(2) an element $\alpha \in \mathfrak{g}$ such that $\alpha \in \mathfrak{g}_0$.

## 2. Manin triples

The following definition is due to Drinfeld [3].

**Definition 2.1.** A Manin triple $(\mathfrak{g},\mathfrak{g}_+,\mathfrak{g}_-)$ consists of a Lie algebra $\mathfrak{g}$, with invariant inner product $(x_1,x_2)$, and isotropic Lie subalgebras $\mathfrak{g}_\pm$ such that $\mathfrak{g} = \mathfrak{g}_+ \oplus \mathfrak{g}_-$.

Given $x \in \mathfrak{g}$, we will denote its projections onto $\mathfrak{g}_\pm$ by $x_\pm$. Denote by $\mathfrak{g}_0$ the subspace of $\mathfrak{g}$ orthogonal to $[\mathfrak{g}_+,\mathfrak{g}_+] \oplus [\mathfrak{g}_-,\mathfrak{g}_-]$.

We will only consider Manin triples such that $\mathfrak{g}$ is finite dimensional and reductive over $\mathbb{C}$. We may associate a Manin triple to the following data:

(1) a simple Lie algebra $\mathbf{k}$, with Borel subalgebra $\mathbf{b}$;
(2) an invariant inner product $(-,-)$ on $\mathbf{k}$;
(3) a parabolic subalgebra $\mathbf{p} \supset \mathbf{b}$.

The parabolic subalgebra $\partial$ has the Levi decomposition $\mathbf{p} = \mathbf{l} \oplus \mathbf{n}$, where $\mathbf{l}$ is reductive and $\mathbf{n}$ is nilpotent. Let $\mathfrak{g}$ be the Lie algebra $\mathfrak{g} = \mathbf{k} \oplus \mathbf{l}$, with inner product
$$(x_1 \oplus y_1, x_2 \oplus y_2) = (x_1,x_2) - (y_1,y_2).$$
The subalgebra
$$\mathfrak{g}_+ = \{x \oplus y \in \mathbf{k} \oplus \mathbf{l} \mid x - y \in \mathbf{n}\} \subset \mathfrak{g}$$
is easily seen to be isotropic, and $\dim \mathfrak{g}_+ = \tfrac{1}{2}\dim\mathfrak{g}$.

Let $\mathfrak{h}$ be the Cartan subalgebra of $\mathbf{k}$ determined by the Borel subalgebra $\mathbf{b}$; if $x \in \mathbf{k}$, denote by $x_\mathfrak{h}$ its projection onto $\mathfrak{h}$. Let $\mathbf{b}_-$ be the Borel subalgebra opposite to $\mathbf{b}$. The subalgebra
$$\mathfrak{g}_- = \{x \oplus y \in (\mathbf{k} \cap \mathbf{b}) \oplus (\mathbf{l} \cap \mathbf{b}_-) \mid (x+y)_\mathfrak{h} = 0\} \subset \mathfrak{g}$$
is isotropic, and together with $\mathfrak{g}_+$ forms a Manin triple.

The two extreme cases of this construction are of especial interest:

(1) If $\mathbf{p} = \mathbf{b}$ is itself a Borel subalgebra, then $\mathfrak{g} = \mathbf{k} \oplus \mathfrak{h}$ and
$$\mathfrak{g}_{\pm} = \{x \oplus h \in \mathbf{b}_{\mp} \oplus \mathfrak{h} \mid x_{\mathfrak{h}} = \pm h\}.$$

This Manin triple will be seen to correspond to the Kazama-Suzuki model.

(2) If $\mathbf{p} = \mathbf{k}$ is all of $\mathbf{k}$, then $\mathfrak{g} = \mathbf{k} \oplus \mathbf{k}$ and
$$\mathfrak{g}_{+} = \{x \oplus x \in \mathbf{k} \oplus \mathbf{k}\},$$
$$\mathfrak{g}_{-} = \{x \oplus y \in \mathbf{b}_{+} \oplus \mathbf{b}_{-} \mid x_{\mathfrak{h}} + y_{\mathfrak{h}} = 0\}.$$

This Manin triple will be seen to correspond to the $G/G$ model (and is also familiar as the Manin triple associated to the classical limit of the quantum group $U_q \mathbf{k}$).

Denote by $\Delta \subset \mathfrak{h}^*$ the roots of $\mathbf{k}$, and by $\tilde{\Delta} \subset \Delta$ the roots of $\mathbf{l} \subset \mathbf{p}$. Let $\Pi \subset \Delta$ be the basis of the root system $\Delta$ determined by $\mathbf{b}$, and let $\Delta = \Delta_{+} \cup \Delta_{-}$ be the associated decomposition of $\Delta$ into positive and negative roots. Then $\mathbf{p}$ has the decomposition into a sum of weight spaces of $\mathbf{k}$:
$$\mathbf{p} = \sum_{\alpha \in \Delta_{+} \cup \tilde{\Delta}} \mathbf{k}_\alpha.$$

To a root $\alpha \in \Delta \subset \mathfrak{h}^*$ is associated a coroot $\alpha^{\vee} \in \Delta^{\vee} \subset \mathfrak{h}$, such that the reflection $s_\alpha$ is given by the formula
$$s_\alpha \beta = \beta - \langle \alpha^{\vee}, \beta \rangle.$$

In particular, $\alpha(\alpha^{\vee}) = 2$. For $\alpha \in \Delta_{+}$, let $x_\alpha \in \mathbf{k}_\alpha$ and $y_\alpha \in \mathbf{k}_{-\alpha}$ be vectors such that
$$[x_\alpha, y_\alpha] = \alpha^{\vee}, \quad [\alpha^{\vee}, x_\alpha] = 2x_\alpha, \quad [\alpha^{\vee}, y_\alpha] = 2y_\alpha.$$

It follows that
$$(x_\alpha, y_\alpha) = \tfrac{1}{2}([\alpha^{\vee}, x_\alpha], y_\alpha) = \tfrac{1}{2}(\alpha^{\vee}, [x_\alpha, y_\alpha]) = \tfrac{1}{2}(\alpha^{\vee}, \alpha^{\vee}).$$

Given the Manin triple $(\mathfrak{g} = \mathbf{k} \oplus \mathbf{l}, \mathfrak{g}_+, \mathfrak{g}_-)$ at level $k$, $\mathfrak{g}_+$ has basis
$$\{x_\alpha \oplus 0 \mid \alpha \in \Delta_+ \setminus \tilde{\Delta}_+\} \cup \{\alpha^{\vee} \oplus \alpha^{\vee} \mid \alpha \in \Pi\} \cup \{x_\alpha \oplus x_\alpha \mid \alpha \in \tilde{\Delta}_+\} \cup \{y_\alpha \oplus y_\alpha \mid \alpha \in \tilde{\Delta}_+\}.$$

If $\{\alpha_i \mid 1 \leq i \leq r\}$ are the simple roots of $\mathbf{k}$, let $\{\omega_i \mid 1 \leq i \leq r\} \subset \mathfrak{h}^*$ be the fundamental weights, characterized by the formula
$$\langle \alpha_i^{\vee}, \omega_j \rangle = \delta_{ij}.$$

We may identify the weights $\omega_i$ with elements of $\mathfrak{h}$ by means of the inner product $(-,-)$ restricted $\mathfrak{h}$. With these notations, $\mathfrak{g}_-$ has dual basis
$$\left\{ \tfrac{2}{(\alpha^{\vee}, \alpha^{\vee})}(y_\alpha \oplus 0) \mid \alpha \in \Delta_+ \right\} \cup \left\{ \tfrac{2}{(\alpha^{\vee}, \alpha^{\vee})}(0 \oplus x_\alpha) \mid \alpha \in \tilde{\Delta}_+ \right\} \cup \left\{ \tfrac{1}{2}(\omega_i \oplus -\omega_i) \mid 1 \leq i \leq r \right\}.$$

Thus, $\mathfrak{g}_0$ has the basis
$$\{\alpha^{\vee} \oplus \alpha^{\vee} \mid \alpha \in \Pi\} \cup \{\omega_i \oplus -\omega_i \mid \alpha_i \in \Pi \cap \tilde{\Delta}\}.$$

If $(\mathfrak{g}, \mathfrak{g}_+, \mathfrak{g}_-)$ is a Manin triple, the inner product on $\mathfrak{g}$ induces an identification of $\mathfrak{g}_-$ with the dual $\mathfrak{g}_+^*$ of $\mathfrak{g}_+$, and vice versa. Denote the adjoint action of $\mathfrak{g}_+$ on itself by $\mathrm{ad}(x)$, and the coadjoint action of $\mathfrak{g}_-$ on $\mathfrak{g}_+$ by $\mathrm{ad}^*(x)$. The choice of a basis $x_i$ of $\mathfrak{g}_+$ determines a dual basis $x^i$ of $\mathfrak{g}_-$, such that the bracket of $\mathfrak{g}$ is given by the formulas

$$[x_i, x_j] = c_{ij}^k x_k,$$
$$[x_i, x^j] = f_i^{jk} x_k + c_{ki}^j x^k,$$
$$[x^i, x^j] = f_k^{ij} x^k.$$

Thus $\mathrm{ad}(x_i) x_j = c_{ij}^k x_k$, while $\mathrm{ad}^*(x^i) x_j = -f_j^{ik} x_k$. The coefficients $c_{ij}^k$ and $f_k^{ij}$ satisfy the identities

$$c_{ij}^k = -c_{ji}^k, \quad f_k^{ij} = -f_k^{ji},$$
$$c_{ij}^m c_{mk}^l + c_{jk}^m c_{mi}^l + c_{ki}^m c_{mj}^l = 0,$$
$$f_m^{ij} f_l^{mk} + f_m^{jk} f_l^{mi} + f_m^{ki} f_l^{mj} = 0,$$
$$c_{mk}^i f_l^{jm} - c_{ml}^i f_k^{jm} - c_{mk}^j f_l^{im} + c_{ml}^j f_k^{im} = c_{kl}^m f_m^{ij}.$$

The first four of these equations say that $\mathfrak{g}_+$ and $\mathfrak{g}_-$ are Lie algebras, while the last equation says that $\mathrm{ad}^*(x)$ is a derivation of the Lie algebra $\mathfrak{g}_+$; we will refer to this as the cocycle formula.

The element $\rho \in \mathfrak{g}$, defined by the formula $\rho = [x_i, x^i]$, has projections onto $\mathfrak{g}_\pm$ equal to

$$\rho_+ = f_j^{ji} x_i \quad \text{and} \quad \rho_- = c_{ij}^j x^i.$$

It is easily seen that $[\rho_+, \rho_-] = 0$, and the cocycle formula shows that $\rho \in \mathfrak{g}_0$.

Let $D : \mathfrak{g}_+ \to \mathfrak{g}_+$ denote the derivation of the Lie algebra $\mathfrak{g}_+$

$$Dx_i = -[\rho, x_i]_+ = (f_k^{jk} c_{ji}^l + c_{jk}^k f_i^{jl}) x_l.$$

**Lemma 2.2.** $Dx_i = f_i^{kl} c_{kl}^j x_j$ and $\mathrm{Tr}(D) = -(\rho, \rho) = -2(\rho_+, \rho_-)$

*Proof.* Let $A : \mathfrak{g}_+ \otimes \mathfrak{g}_- \to \mathbb{R}$ be the bilinear form $A(x, y) = -\mathrm{Tr}(\mathrm{ad}(x)\,\mathrm{ad}^*(y))$, with matrix $A_i^j = c_{il}^k f_k^{jl}$. Taking the trace in the cocycle formula over the indices $i$ and $l$, we see that

$$-A_k^j - c_{mi}^i f_k^{jm} - c_{mk}^j f_i^{im} - D_k^j + A_k^j = 0.$$

The first and last terms cancel, and we obtain the formula for $Dx_i$. Taking the trace over the indices $j$ and $k$, we obtain the formula for $\mathrm{Tr}(D)$. $\square$

Denote by $\langle -, - \rangle$ the Killing form of $\mathfrak{g}$, and by $\langle -, - \rangle_\pm$ the Killing forms of $\mathfrak{g}_\pm$. The following formulas are straightforward:

$$\langle x_i, x_j \rangle = 2 \langle x_i, x_j \rangle_+, \quad \langle x_i, x^j \rangle = -D_i^j - 2A_i^j, \quad \langle x^i, x^j \rangle = 2 \langle x^i, x^j \rangle_-.$$

From these formulas, we see that $\langle x_i, x^i \rangle = -3\,\mathrm{Tr}(D)$ and $\mathrm{Tr}(D^2) = \tfrac{1}{2} \langle \rho, \rho \rangle$.

Let us illustrate these formulas in the case of the Manin triple $(\mathfrak{g} = \mathbf{k} \oplus \mathbf{l}, \mathfrak{g}_+, \mathfrak{g}_-)$. From the explicit basis introduced above for $\mathfrak{g} = \mathbf{k} \oplus \mathbf{l}$, it follows that

$$\rho = \sum_{\alpha \in \Delta_+} \frac{2\alpha^{\vee}}{(\alpha^{\vee}, \alpha^{\vee})} \oplus \sum_{\alpha \in \tilde{\Delta}_+} \frac{2\alpha^{\vee}}{(\alpha^{\vee}, \alpha^{\vee})} = \frac{2}{k+h} \rho(\mathbf{k}) \oplus \rho(\mathbf{l}),$$

where

$$\rho(\mathbf{k}) = \frac{1}{2} \sum_{\alpha \in \Delta_+} \alpha \quad \text{and} \quad \rho(\mathbf{l}) = \frac{1}{2} \sum_{\alpha \in \tilde{\Delta}_+} \alpha,$$

with elements of $\mathfrak{h}$, by means of its inner product $(-,-)$. Hence, we see that

$$(k+h)\rho_{\pm} = (\rho(\mathbf{k}) \pm \rho(\mathbf{l})) \oplus (\pm \rho(\mathbf{k}) + \rho(\mathbf{l})),$$

and that

$$(k+h) \operatorname{Tr}(D) = -2(\rho(\mathbf{k}), \rho(\mathbf{k})) + 2(\rho(\mathbf{l}), \rho(\mathbf{l})).$$

Take as an example $\mathfrak{g} = \mathbf{k} \oplus \mathfrak{h}$ at level $k$. From Freudenthal's formula

$$\frac{(\rho, \rho)}{(\theta, \theta)} = \frac{h \dim \mathbf{k}}{24},$$

where $\theta$ is the highest root of $\mathbf{k}$, we see that

$$\operatorname{Tr}(D) = -\frac{h \dim \mathbf{k}}{6(k+h)}.$$

## 3. A family of $N = 2$ superconformal field theories

Let $(\mathfrak{g}, \mathfrak{g}_+, \mathfrak{g}_-)$ be a Manin triple, with dual bases $x_i$ of $\mathfrak{g}_+$ and $x^i$ of $\mathfrak{g}_-$. We may rewrite the $N = 1$ superconformal model associated to $\mathfrak{g}$ in this basis. The Kac-Moody currents $\{J_i(z), J^i(z)\}$ have operator products

$$J_i(z) \cdot J_j(w) = -\frac{\frac{1}{2}\langle x_i, x_j \rangle}{(z-w)^2} + \frac{c_{ij}^k J_k(w)}{z-w},$$

$$J_i(z) \cdot J^j(w) = \frac{\delta_i^j - \frac{1}{2}\langle x_i, x^j \rangle}{(z-w)^2} + \frac{f_i^{jk} J_k(w) + c_{ki}^j J^k(w)}{z-w},$$

$$J^i(z) \cdot J^j(w) = -\frac{\frac{1}{2}\langle x^i, x^j \rangle}{(z-w)^2} + \frac{f_k^{ij} J^k(w)}{z-w},$$

while the free fermionic fields $a_i(z)$ and $a^i(z)$ have operator products

$$a_i(z) \cdot a^j(w) \sim \frac{\delta_i^j}{z-w},$$

$$a_i(z) \cdot a_j(w) \sim a^i(z) \cdot a^j(w) \sim 0.$$

The composite currents $I_i(z)$ and $I^i(z)$ are given by the formulas

$$I_i = J_i - c_{ij}^k a^j a_k - \tfrac{1}{2} f_i^{jk} a_j a_k,$$

$$I^i = J^i - f_k^{ij} a_j a^k - \tfrac{1}{2} c_{jk}^i a^j a^k.$$

Now suppose that $\alpha \in \mathfrak{g}_0$. Writing $\alpha_+ = p^i x_i$ and $\alpha_- = q_i x^i$, the condition that $\alpha \in \mathfrak{g}_0$ is equivalent to the formulas

$$p_i c^i_{jk} = q^i f_i^{jk} = 0.$$

Thus, the fields $(\alpha_\pm, I)$ are given by the formulas

$$(\alpha_+, I) = p^i J_i - p^i c^k_{ij} a^j a_k, \quad (\alpha_-, I) = q_i J^i - q_i f^{ij}_k a_j a^k.$$

Define fermionic fields

$$\mathbb{G}^+ = J_i a^i - \tfrac{1}{2} c^k_{ij} a^i a^j a_k + (\alpha_-, \partial a),$$
$$\mathbb{G}^- = J^i a_i - \tfrac{1}{2} f^{ij}_k a_i a_j a^k + (\alpha_+, \partial a),$$

where the fields $(\alpha_\pm, \partial a)$ are given by the formulas

$$(\alpha_+, \partial a) = p^i \partial a_i, \quad (\alpha_-, \partial a) = q_i \partial a^i.$$

It is not difficult to verify that $\mathbb{G}^\pm(z) \cdot \mathbb{G}^\pm(w) \sim 0$: when $\alpha_\pm = 0$, this follows from the form of the central extension of the fields $J_i(z)$ and $J^i(z)$, while the terms $(\alpha_\pm, \partial a)$ do not contribute to the operator products by the hypothesis that $\alpha \in \mathfrak{g}_0$. We now calculate the operator product $\mathbb{G}^+(z) \cdot \mathbb{G}^-(w)$.

**Lemma 3.1.**

$$\mathbb{G}^+(z) \cdot \mathbb{G}^-(w) \sim \frac{\tfrac{1}{2} \dim \mathfrak{g} - (\rho, \rho)}{(z-w)^3} + \frac{a^i(w) a_i(w) + (\rho, I(w))}{(z-w)^2}$$
$$+ \frac{\tfrac{1}{2}(J(w), J(w)) + \partial a^i(w) a_i(w) + \tfrac{1}{2}(\rho, \partial I(w))}{z-w}$$
$$- \frac{(\alpha, \alpha)}{(z-w)^3} + \frac{(\alpha_+ - \alpha_-, I(w))}{(z-w)^2} + \frac{(\alpha_+, \partial I(w))}{z-w}.$$

*Proof.* Denote by $\mathbb{G}^\pm_0(z)$ the corresponding fields with $\alpha = 0$. We first prove the lemma with $\alpha = 0$, and then for arbitrary $\alpha$. First, we calculate that

$$J_i(z) a^i(z) \cdot J^j(w) a_j(w) \sim \frac{A_3}{(z-w)^3} + \frac{A_2}{(z-w)^2} + \frac{A_1 + B_1}{(z-w)^2},$$

where

$$A_3 = \tfrac{1}{2}(\dim \mathfrak{g} - \langle x_i, x^i \rangle) = \tfrac{1}{2} \dim \mathfrak{g} + \tfrac{3}{2} \operatorname{Tr}(D),$$
$$A_2 = (\delta^j_i - \tfrac{1}{2}\langle x_i, x^j \rangle) a^i(w) a_j(w) + (\rho, J(w)),$$
$$= a^i(w) a_i(w) + (-\tfrac{1}{2} D^j_i + A^j_i) a^i(w) a_j(w) + (\rho, I(w)),$$
$$A_1 = (\delta^j_i - \tfrac{1}{2}\langle x_i, x^j \rangle) \partial a^i(w) a_j(w) + J_i(w) J^i(w),$$
$$= \partial a^i(w) a_i(w) - \tfrac{1}{2} D^j_i a^i(w) \partial a_j(w) + A^j_i \partial a^i(w) a_j(w)$$
$$+ \tfrac{1}{2}\big(J_i(w) J^i(w) + J^i(w) J_i(w) + (\rho, \partial I(w))\big),$$
$$B_1 = f^{jk}_i J_k(w) a^i(w) a_j(w) + c^j_{ki} J^k(w) a^i(w) a_j(w).$$

Here, we have made use of the formulas
$$J_i(z)J^i(z) = \tfrac{1}{2}\big((J(z), J(z)) + (\rho, \partial J(z))\big)$$
and
$$(\rho, \partial I(z)) = (\rho, \partial J(z)) + D_i^j a^i(z) a_j(z).$$

The term $B_1$ is cancelled by the operator products
$$J_i(z)a^i(z) \cdot (-\tfrac{1}{2}f_l^{jk} a_j(w) a_k(w) a^l(w)) \sim -\frac{f_i^{jk} J_k(w) a^i(w) a_j(w)}{z - w},$$
$$(-\tfrac{1}{2}c_{jk}^l a^j(z) a^k(z) a_l(z)) \cdot J^i(w) a_i(w) \sim -\frac{c_{kj}^i J^k(w) a^i(w) a_j(w)}{z - w}.$$

The calculation of the operator product $\mathbb{G}_0^+(z) \cdot \mathbb{G}_0^-(w)$ is completed by the formula
$$(-\tfrac{1}{2}c_{ij}^k a^i(z) a^j(z) a_k(z)) \cdot (-\tfrac{1}{2}f_r^{pq} a_p(w) a_q(w) a^r(w)) \sim$$
$$-\frac{\tfrac{1}{2}\operatorname{Tr}(D)}{(z-w)^3} + \frac{(\tfrac{1}{2}D_i^j - A_i^j)a^i(w)a_j(w)}{(z-w)^2} + \frac{\tfrac{1}{2}D_i^j a^i(w)\partial a_j(w)}{z-w} - \frac{A_i^j \partial a^i(w) a_j(w)}{z-w}.$$

To extend the result to general $\alpha_\pm$, we use the formulas
$$\mathbb{G}_0^+(z) \cdot (\alpha_+, \partial a(w)) \sim \frac{(\alpha_+, I(w))}{(z-w)^2} + \frac{(\alpha_+, \partial I(w))}{z-w},$$
$$(\alpha_-, \partial a(z)) \cdot \mathbb{G}_0^-(w) \sim -\frac{(\alpha_-, I(w))}{(z-w)^2},$$
$$(\alpha_-, \partial a(z)) \cdot (\alpha_+, \partial a(w)) \sim -\frac{(\alpha, \alpha)}{(z-w)^3}. \quad \square$$

Motivated by this lemma, we introduce fields
$$\mathbb{J} = a^i a_i + (\rho, I) + (\alpha_+ - \alpha_-, I),$$
$$\mathbb{T} = \tfrac{1}{2}\big((J, J) + (\partial a, a) - (a, \partial a) + (\alpha, \partial I)\big).$$

Then the above lemma shows that
$$\mathbb{G}^+(z) \cdot \mathbb{G}^-(w) \sim \frac{\tfrac{1}{2}d}{(z-w)^3} + \frac{\mathbb{J}(w)}{(z-w)^2} + \frac{\mathbb{T}(w) + \tfrac{1}{2}\mathbb{J}(w)}{z-w},$$
with $d = \tfrac{1}{2}\dim \mathfrak{g} - (\rho, \rho) - (\alpha, \alpha)$.

There is an especially natural choice for $\alpha$, namely $\alpha = -\rho_+ + \rho_-$. With this value of $\alpha$, the field $\mathbb{J}(z)$ takes the especially simple form $\mathbb{J} = a^i a_i$, while the central charge becomes simply $d = \tfrac{1}{2}\dim \mathfrak{g}$.

To verify the remainder of the operator products of the $N = 2$ chiral algebra, we use the following lemma (which has been found independently by Figueroa-O'Farill [4]).

**Lemma 3.2.** *Let $G^\pm(z)$ be fermionic fields such that $G^\pm(z) \cdot G^\pm(w) \sim 0$,*

$$G^+(z) \cdot G^-(w) \sim \frac{d}{(z-w)^3} + \frac{J(w)}{(z-w)^2} + \frac{T(w) + \frac{1}{2}\partial J(w)}{z-w},$$

*for some scalar $d$ and bosonic fields $J(z)$ and $T(z)$, and*

$$J(z) \cdot G^\pm(w) \sim \pm \frac{G^\pm(w)}{z-w}.$$

*Then these fields satisfy the operator products of the $N = 2$ superconformal algebra.*

*Proof.* The proof is a straighforward exercise in the application of the formulas of the appendix. We will use the formulas

$$T = [G^+G^-]_1 - \tfrac{1}{2}\partial J = [G^-G^+]_1 + \tfrac{1}{2}\partial J,$$
$$J = [G^+G^-]_2 = -[G^-G^+]_2,$$
$$d = [G^+G^-]_3 = [G^-G^+]_3.$$

(1) We start by calculating $J(z) \cdot J(w)$:

$$[JJ]_n = [J[G^+G^-]_2]_n = \sum_{i=0}^{n-1} \binom{n-1}{i}[[JG^+]_{n-i}G^-]_{i+2} + [G^+[JG^-]_n]_2$$

$$= \begin{cases} [[JG^+]_1 G^-]_2 + [G^+[JG^-]_1]_2 = 0, & n = 1, \\ [[JG^+]_1 G^-]_3 = d, & n = 2, \\ [[JG^+]_1 G^-]_{n+1} = 0, & n > 2. \end{cases}$$

(2) Next we calculate $J(z) \cdot T(w)$:

$$[JT]_n = \tfrac{1}{2}([J[G^+G^-]_1]_n + [J[G^-G^+]_1]_n)$$
$$= \tfrac{1}{2}\sum_{i=0}^{n-1}\binom{n-1}{i}([[JG^+]_{n-i}G^-]_{i+1} + [[JG^-]_{n-i}G^+]_{i+1})$$
$$+ \tfrac{1}{2}([G^+[JG^-]_n]_1 + [G^-[JG^+]_n]_1)$$

$$= \begin{cases} \tfrac{1}{2}([[JG^+]_1 G^-]_1 + [[JG^-]_1 G^+]_1 + [G^+[JG^-]_1]_1 + [G^-[JG^+]_1]_1) = 0, & n = 1, \\ \tfrac{1}{2}([[JG^+]_1 G^-]_2 + [[JG^-]_1 G^+]_2) = J, & n = 2, \\ \tfrac{1}{2}([[JG^+]_1 G^-]_3 + [[JG^-]_1 G^+]_3) = d - d = 0, & n = 3, \\ \tfrac{1}{2}([[JG^+]_1 G^-]_n + [[JG^-]_1 G^+]_n) = 0, & n > 3. \end{cases}$$

(3) Next we calculate $G^+(z) \cdot T(w)$. Using the formula $T = [G^+G^-]_1 - \tfrac{1}{2}\partial J$, we see that

$$[G^+T]_n = [G^+[G^+G^-]_1]_n - \tfrac{1}{2}[G^+(\partial J)]_n = -[G^+[G^+G^-]_n]_1 - \tfrac{1}{2}[G^+(\partial J)]_n.$$

When $n = 1$, this shows that $[G^+T]_1 = -[G^+T]_1 - [G^+(\partial J)]_1$, so that

$$[G^+T]_1 = -\tfrac{1}{2}[G^+(\partial J)]_1 = -\tfrac{1}{2}\partial[G^+J]_1 = \tfrac{1}{2}\partial G^+.$$

For $n = 2$, we have
$$[G^+T]_2 = -[G^+J]_1 - \tfrac{1}{2}\partial[G^+J]_2 - \tfrac{1}{2}[G^+J]_1 = \tfrac{3}{2}G^+.$$

It is easily seen that $[G^+T]_n = 0$ for $n > 2$. The calculation of $G^-(z) \cdot T(w)$ is similar, except that we use the formula $T = [G^-G^+]_1 + \tfrac{1}{2}\partial J$.

(5) Finally, we calculate $T(z) \cdot T(w)$:
$$\begin{aligned}[TT]_n &= \tfrac{1}{2}([T[G^+G^-]_1]_n + [T[G^-G^+]_1]_n) \\ &= \tfrac{1}{2}\sum_{i=0}^{n-1}\binom{n-1}{i}([[TG^+]_{n-i}G^-]_{i+1} + [[TG^-]_{n-i}G^+]_{i+1}) \\ &\quad + \tfrac{1}{2}([G^+[TG^-]_n]_1 + [G^-[TG^+]_n]_1).\end{aligned}$$

For $n = 1$, this equals
$$\tfrac{1}{2}\Big([(\partial G^+)G^-]_1 + [(\partial G^-)G^+]_1 + [G^+(\partial G^-)]_1 + [G^-(\partial G^+)]_1\Big)$$
$$= \tfrac{1}{2}(\partial[G^+G^-]_1 + \partial[G^-G^+]_1) = \partial T.$$

For $n = 2$, we obtain
$$\tfrac{1}{2}\Big([[TG^+]_2G^-]_1 + [[TG^-]_2G^+]_1 + [[TG^+]_1G^-]_2 + [[TG^-]_1G^+]_2$$
$$+ [G^+[TG^-]_2]_1 + [G^-[TG^+]_2]_1\Big) = 2T.$$

For $n = 3$, we obtain
$$\tfrac{1}{2}\Big(2[[TG^+]_2G^-]_2 + 2[[TG^-]_2G^+]_2 + [[TG^+]_1G^-]_3 + [[TG^-]_1G^+]_3\Big) = 0.$$

Finally, for $n = 4$ we obtain
$$\tfrac{1}{2}\Big(3[[TG^+]_2G^-]_3 + 3[[TG^-]_2G^+]_2 + [[TG^+]_1G^-]_4 + [[TG^-]_1G^+]_4\Big) = \tfrac{3}{2}d. \quad \square$$

To apply this lemma to the fields $\mathbb{G}^\pm(z)$, we must calculate the operator products $\mathbb{J}(z) \cdot \mathbb{G}^\pm(w)$. Observe that
$$a^i(z)a_i(z) \cdot \mathbb{G}^\pm(w) \sim -\frac{(\rho_\mp, a(w))}{(z-w)^2} \pm \frac{(\alpha_\mp, a(w))}{(z-w)^2} \pm \frac{\mathbb{G}^\pm(w)}{z-w}.$$

Now, if $\alpha \in \mathfrak{g}_0$, we may also prove the formula
$$(\alpha, I(z)) \cdot \mathbb{G}^\pm(w) \sim \frac{(\alpha_\mp, a(w))}{(z-w)^2}.$$

The operator products $\mathbb{J}(z) \cdot \mathbb{G}^\pm(w) \sim \pm\mathbb{G}^\pm(w)/(z-w)$ follow immediately.

Until now, we have not payed attention to the boundary conditions of the $N = 2$ superconformal algebra. In fact, there is a whole family of natural choices, parametrized by an angle $\theta$, for which the fields $\mathbb{G}^\pm(z)$ satisfy the conditions
$$\mathbb{G}^\pm(e^{2\pi i}z) = e^{\pm i\theta}\mathbb{G}^\pm(z).$$

Of these, only the cases where $e^{i\theta} = \pm 1$ have an underlying $N = 1$ superconformal symmetry: the choice of periodic boundary conditions $e^{i\theta} = 1$ goes by the name of the Ramond sector, while the choice of anti-periodic boundary conditions $e^{i\theta} = -1$ is called the Neveu-Schwarz sector. We may realize the $N = 2$ superconformal algebra with these boundary conditions in the model of this section by taking the fermions to have boundary conditions

$$a^k(e^{2\pi i}z) = e^{i\theta}a^k(z) \quad , \quad a_k(e^{2\pi i}z) = e^{i\theta}a_k(z).$$

The twisting construction of Witten and Eguchi-Yang is a way of constructing a topological field theory in its Ramond (Neveu-Schwarz) sector from an $N = 2$ theory in its Neveu-Schwarz (Ramond) sector. The Hilbert spaces of the two theories, are identical, and all that is changed is the stress-energy tensor $\mathbb{T}(z)$, which is replaced by its twist $\mathbb{T}_{\text{top}}(z) = \mathbb{T}(z) + \frac{1}{2}\partial\mathbb{J}(z)$. This has the effect of modifying the conformal dimensions of the fields in the theory, subtracting from them half of their U(1) charge. In particular, the field $\mathbb{G}^+(z)$ now has conformal dimension 1. In the Ramond sector of the topological theory, this allows us to think of the zero mode $\mathbb{Q}$ of the field $\mathbb{G}^+$ as a differential on the space of states of the theory, graded by the zero mode of the field $\mathbb{J}$. (The operator $\mathbb{Q}$ may be identified with the mode $\mathbb{G}^+_{1/2}$ of the untwisted $N = 2$ theory.)

The operator $\mathbb{Q}$ raises degree by 1, and its cohomology is by definition the space of physical states of the topological field theory. By the formula

$$\mathbb{T}_{\text{top}}(z) = [\mathbb{Q}, \mathbb{G}^-(z)],$$

we see that the Virasoro algebra acts trivially on this cohomology space, explaining why it is called a topological field theory. In our model, this differential

$$\mathbb{Q} = \text{Res}\Big(J_i a^i - \tfrac{1}{2}c^k_{ij}a^i a^j a_k\Big)$$

calculates the semi-infinite cohomology for the projective action of the affine Lie algebra $\widehat{\mathfrak{g}_+}$ acting through the currents $J_i(z)$. Note that this differential is independent of $\alpha \in \mathfrak{g}_0$, and hence so is the physical Hilbert space. By contrast, the zero mode of $\mathbb{J}$, which defines the grading, and the zero-mode $\Delta$ of the spin 2 field $\mathbb{G}^-$ do depend on $\alpha$. The operator $\Delta$ is important in topological field theory, since it expresses the coupling of the theory to topological gravity, and in particular is needed in the definition of the equivariant cohomology of the topological field theory [5]. It is natural to conjecture that the equivariant cohomology of the model is independent of $\alpha$.

## Appendix. The operator product expansion and vertex algebras

In this paper, we have made free use of the operator product expansion of conformal field theory, as axiomatized in the notion of a vertex algebra. A vertex algebra is a $\mathbb{Z}/2$-graded vector space $V$, together with an even map $\partial : V \to V$ (which may be interpreted as differentiation), and a sequence of bilinear products

$$[AB]_n : V \otimes V \to V, \quad n \in \mathbb{Z},$$

such that for given $A$ and $B$, the product $[AB]_n$ vanishes for $n$ sufficiently large. These products are the Laurent coefficients of the operator product of the two fields in two-dimensional conformal field theory (Belavin-Polyakov-Zamolodchikov [1]):

$$A(z) \cdot B(w) = \sum_{n=-\infty}^{\infty} \frac{[AB]_n(w)}{(z-w)^n}.$$

For this reason, elements of a vertex algebra are sometimes referred to as fields. The product $[AB]_0$ is called the normal product of $A$ and $B$, and should be thought of as a renormalized product on $V$.

These products satisfy axioms, first written down by Borcherds [2], which are reminiscent of those of those of a Lie algebra. (We have reindexed his products in order to agree with the conventions of physicists: thus, we write $[AB]_n$ where Borcherds writes $A_{n-1}B$.)

**(Jacobi)**

$$[[AB]_m C]_n = \sum_{i=0}^{\infty} (-1)^i \binom{m-1}{i} \Big( [A[BC]_{n+i}]_{m-i} + (-1)^{m+|A||B|} [B[AC]_{i+1}]_{m+n-i-1} \Big)$$

Here $\binom{a}{i} = a(a-1)\ldots(a-i+1)/i!$.

**(Commutativity)**

$$[BA]_n = (-1)^{n+|A||B|} \sum_{i=0}^{\infty} \frac{(-1)^i}{i!} \partial^i [AB]_{n+i}$$

**(Identity)** There is an even element 1 such that $\partial 1 = 0$ and for all $A \in V$,

$$[1A]_n = \begin{cases} A, & n = 0, \\ 0, & n \neq 0. \end{cases}$$

For $m = n = 0$, the Jacobi rule simply says that

$$[[AB]_0 C]_n = \sum_{i \leq 0} [A[BC]_{-i}]_i + \sum_{i > 0} (-1)^{|A||B|} [B[AC]_i]_{-i}.$$

Subtracting the corresponding formula for $(-1)^{|A||B|}[[BA]_0 C]_0$, we see that

$$[[AB]_0 C]_n - (-1)^{|A||B|} [[BA]_0 C]_0 = [A[BC]_0]_0 - (-1)^{|A||B|} [B[AC]_0]_0.$$

The above axioms imply the following properties of the derivation $\partial$.

**Proposition A.1.** $[(\partial A)B]_n = (1-n)[AB]_{n-1}$ and $\partial[AB]_n = [(\partial A)B]_n + [A(\partial B)]_n$

*Proof.* To calculate $[(\partial A)B]_n$, we use the Jacobi rule, and the formula $\partial A = [A1]_{-1}$:

$$[(\partial A)B]_n = [[A1]_{-1}B]_n = \sum_{i=0}^{\infty}(-1)^i\binom{-2}{i}\Big([A[1B]_{n+i}]_{-i-1} - [1[AB]_{i+1}]_{n-i-2}\Big).$$

Note that $(-1)^i\binom{-2}{i} = i+1$. If $n < 1$, only the first term on the left-hand side contributes, with $i = -n$. On the other hand, if $n > 1$, only the second term contributes, with $i = n - 2$.

To calculate $[A(\partial B)]_n$, we apply the commutativity axiom:

$$\begin{aligned}
[A(\partial B)]_n &= \sum_{i=0}^{\infty} \frac{(-1)^{n+i+|A||B|}}{i!} \partial^i[(\partial B)A]_{n+i} \\
&= \sum_{i=0}^{\infty} \frac{(-1)^{n+i+|A||B|}}{i!}(1-n-i)\partial^i[BA]_{n+i-1} \\
&= (n-1)\sum_{i=0}^{\infty} \frac{(-1)^{(n-1)+i+|A||B|}}{i!}\partial^i[BA]_{n+i-1} + \sum_{i=0}^{\infty} \frac{(-1)^{n+i+|A||B|}}{i!}\partial^i[BA]_{n+i} \\
&= (n-1)[AB]_{n-1} + \partial[AB]_n. \quad \square
\end{aligned}$$

In this paper, we will only use the following consequence of the Jacobi rule, discovered by Sevrin et al. [9]. This formula shows the extent to which the map $B \mapsto [AB]_m$, $m > 0$, behaves like a derivation.

**Proposition A.2.** *If $m > 0$,*

$$\begin{aligned}
[A[BC]_n]_m &= \sum_{i=0}^{m-1}\binom{m-1}{i}[[AB]_{m-i}C]_{n+i} + (-1)^{|A||B|}[B[AC]_m]_n \\
&= \sum_{i=0}^{\infty}\frac{(-1)^i}{i!}[[(\partial^i A)B]_m C]_{n+i} + (-1)^{|A||B|}[B[AC]_m]_n.
\end{aligned}$$

*Proof.* If $m > 0$, the Jacobi rule may be rewritten

$$[[AB]_m C]_n = \sum_{i=0}^{m-1}(-1)^i\binom{m-1}{i}\Big([A[BC]_{n+i}]_{m-i} - (-1)^{|A||B|}[B[AC]_{m-i}]_{n+i}\Big).$$

It follows that

$$\sum_{j=0}^{m-1}\binom{m-1}{j}[[AB]_{m-j}C]_{n+j} = \sum_{j=0}^{m-1}\sum_{i=0}^{m-j-1}(-1)^i\binom{m-1}{j}\binom{m-j-1}{i}$$
$$\Big([A[BC]_{n+i+j}]_{m-i-j} - (-1)^{|A||B|}[B[AC]_{m-i-j}]_{n+i+j}\Big).$$

The proposition now follows from the combinatorial identity

$$\sum_{\{0\leq i,j\leq m-1 | i+j=k\}}(-1)^i\binom{m-1}{j}\binom{m-j-1}{i} = \begin{cases}1, & k=0, \\ 0, & k>0,\end{cases}$$

which is a trivial consequence of the identity $(x+(1-x))^{m-1} = 1$. $\square$

The singular part of the operator product expansion is denoted

$$A(z) \cdot B(w) \sim \sum_{n>0} \frac{[AB]_n(w)}{(z-w)^n}.$$

We will call the algebraic structure similar to a vertex algebra, but with products $[AB]_n$ only for $n > 0$, satisfying the commutativity and identity axioms as well as the formula of Propositions A.2 and A.1, a chiral algebra. The reader is warned that this is not a standard piece of terminology, but it will be useful to have a word for this notion.

A stress-energy tensor in a vertex algebra is an even element $T$ such that for all $A \in V$,

$$[TA]_1 = \partial A,$$

and for some scalar $c$ (the central charge of $T$),

$$T(z) \cdot T(w) \sim \frac{\frac{1}{2}c \cdot 1(w)}{(z-w)^4} + \frac{2T(w)}{(z-w)^2} + \frac{\partial T(w)}{z-w}.$$

It is usual to denote the scalar multiples $c \cdot 1(w)$ of the identity 1 simply by $c$. An element $A$ of $V$ has conformal dimension $a \in \mathbb{R}$ if $[TA]_2 = aA$. For example, the conformal dimension of 1 is zero, while the conformal dimension of $T$ is 2.

**Lemma A.3.** *The conformal dimension of $[AB]_0$ is the sum of the conformal dimensions of $A$ and $B$.*

*Proof.* By the Leibniz rule, we see that

$$[T[AB]_0]_2 = [[TA]_2 B]_0 + [[TA]_1 B]_1 + [A[TB]_2]_0 = (a+b)[AB]_0 + [(\partial A)B]_1.$$

The second term vanishes by Proposition A.1. □

To a chiral algebra $V$ (and in particular to a vertex algebra) is associated a Lie algebra

$$L(V) = V[z, z^{-1}]/\{(\partial A)z^n + nAz^{n-1}\},$$

with bracket

$$[Az^n, Bz^m] = \sum_{k \geq 0} \binom{n}{k} [AB]_{k+1} z^{n+m-k}.$$

If $A$ has conformal dimension $a$, one denotes the element $Az^n$ of the Lie algebra $L(V)$ by $A_{n-a+1}$. It follows from the properties of the identity that $1_n = 0$ for $n \neq 0$, and that $1_0$ lies in the centre of $L(V)$. The modes of the stress-energy tensor satisfy the relations of the Virasoro algebra,

$$[T_n, T_m] = (n-m)T_{n+m} + \frac{c}{12}n(n^2-1)1_{n+m}.$$

By a representation of a chiral algebra, we will mean a representation of its associated Lie algebra $L(V)$. We only consider representations in the category $\mathcal{O}$, the definition of which is analogous to that of the category $\mathcal{O}$ for an affine Kac-Moody algebra or the Virasoro algebra.

An element $A$ is a primary field of conformal dimension $a$ if
$$T(z) \cdot A(w) \sim \frac{aA(w)}{(z-w)^2} + \frac{\partial A(w)}{z-w}.$$

The modes of a primary field $A$ in $L(V)$ satisfy the following commutation relations with the Virasoro algebra spanned by $T_n$:
$$[T_n, A_m] = ((a-1)n - m)A_{n+m}.$$

## References

1. A.A. Belavin, A.M. Polyakov and A. M. Zamolodchikov, *Infinite conformal symmetry in two-dimensional quantum field theories,* Nucl. Phys. **B241** (1984), 333–380.
2. R. E. Borcherds, *Vertex operator algebras, Kac-Moody algebras and the monster,* Proc. Natl. Acad. Sci. USA **83** (1986), 3068–3071.
3. V.G. Drinfeld, *Quantum groups*, Proc. Int. Cong. Math., Berkeley, Calif., 1986. I, pp. 798–820.
4. J.M. Figueroa-O'Farill, *Affine algebra, N = 2 superconformal algebras, and gauged WZNW models*, hep-th/9306164.
5. E. Getzler, *Two-dimensional topological gravity and equivariant cohomology*, preprint hep/th-9305013.
6. V.G. Kac and I.T. Todorov, *Superconformal current algebras and their unitary representations*, Commun. Math. Phys. **102** (1985), 337–347.
7. Y. Kazama and H. Suzuki, *New N = 2 superconformal field theories and superstring compactification,* Nucl. Phys. **B321** (1989), 232–268.
8. S. Parkhomenko, *Extended superconformal current algebras and finite-dimensional Manin triples*, Sov. Phys. JETP, **75** (1992), 1–3.
9. A. Sevrin, W. Troust, A. van Proyen and Ph. Spindel, *Extended supersymmetric σ-models on group manifolds (II). Current algebras.* Nucl. Phys. **B211** (1988/89), 465–492.
10. Ph. Spindel, A. Sevrin, W. Troust and A. van Proyen, *Extended supersymmetric σ-models on group manifolds (I). The complex structures.* Nucl. Phys. **B308** (1988), 662-698.
11. K. Thielemans, *A Mathematica package for computing operator product expansions,* Int. Jour. Mod. Physics **C2** (1991), 787–798.

Department of Mathematics, MIT, Cambridge MA 02139 USA
*E-mail address*: getzler@math.mit.edu